\documentclass[aps,prl,twocolumn,showpacs,superscriptaddress]{revtex4-1}
\usepackage{graphicx,color}
\usepackage{amsmath}
\usepackage{amssymb}
\usepackage{bm}

\renewcommand{\Re}{\mathop{\text{Re}}\nolimits}
\renewcommand{\Im}{\mathop{\text{Im}}\nolimits}

\newcommand{\ket}[1]{|{#1}\rangle}
\newcommand{\bra}[1]{\langle{#1}|}

\newcommand{\bracket}[2]{\langle#1|#2\rangle}

\definecolor{dgreen}{rgb}{0,0.5,0}

\definecolor{delete}{cmyk}{0.5,0,0,0}

\begin{document}

\title{Spontaneous Symmetry Breaking via Measurement:\\
From Bose-Einstein Condensates to Josephson Effect}



\author{Takaaki Monnai}
\affiliation{Waseda Institute for Advanced Study, Waseda University, Tokyo 169-8050, Japan}
\author{Mauro Iazzi}
\affiliation{Theoretische Physik, ETH Zurich, 8093 Zurich, Switzerland}
\author{Kazuya Yuasa}
\affiliation{Department of Physics, Waseda University, Tokyo 169-8555, Japan}




\begin{abstract}
Why does spontaneous symmetry breaking occur?
Why is a state breaking symmetry realized?
We explore an idea that measurement selects such a state even if a system is given in a state respecting the symmetry of the system.
We point out that the spectrum of the relevant observable is important, and simply apply the projection postulate for quantum measurement.
We first show that this approach correctly describes the well-known interference of Bose-Einstein condensates. We then examine a fermionic system and prove that superconducting states with a definite relative phase are selected by the measurement of the current flowing between two superconductors, eliminating the need to assume the presence of an \textit{a priori} phase to explain the Josephson effect.
\end{abstract}


\maketitle

Spontaneous symmetry breaking plays important roles in various areas in physics, from condensed-matter physics to elementary-particle physics.
It is the symmetry breaking that makes possible a variety of rich and intriguing phenomena in nature out of such simple and elegant fundamental laws built on the principles of symmetry \cite{ref:MoreIsDifferent}.
Why, however, do macroscopic systems tend to break symmetry?
Why are the states with broken symmetry favored and realized?
The importance of the cluster property is stressed and its relevance to the stability of quantum states is discussed \cite{ref:QFT-Weinberg2Chap19,ref:MiyaderaShimizuPRL89}.
In this Letter, we explore an idea that \textit{measurement} selects such a state. 
We show that a state breaking symmetry is realized as an eigenstate of a relevant observable, to which the system is projected by the measurement of the observable.

A typical system exhibiting spontaneous symmetry breaking is Bose-Einstein condensate (BEC) \cite{BEC-Stringari}. 
When two independently prepared BECs overlap, an interference pattern is observed \cite{ref:InterferenceBEC}. 
This is possible since the $\text{U}(1)$ symmetry of the system is broken so that the relative phase between the condensates is well defined.
On the other hand, the following question is asked: what if each condensate consists of exactly $N$ atoms?
Since the number is fixed, the phase is uncertain, due to the phase-number uncertainty relation.
The condensates therefore would not interfere with each other.
This dilemma can be resolved by the idea of ``measurement-induced interference'' \cite{JavanainenCiracWalls,ref:CastinDalibard,ReviewsBEC,ref:LeggettQuantumLiquids,ref:BECPethickSmith}: even if the phases of the condensates are uncertain, detecting the atoms in the overlapping condensates gradually builds up a relative phase, and an interference pattern is observed in each \textit{single} experiment.
This idea triggered intensive studies on the (related) subjects, both theoretical  \cite{IntTh,Boston} and experimental \cite{Hadzibabic,Schmidmayer}.

In this Letter, we show that the measurement-induced interference can be explained simply by the projection postulate in quantum mechanics \cite{ref:LeggettSols,ref:LeggettQuantumLiquids}. 
It is important to identify 
the observable to be measured and its spectrum.
For the interference of BECs, the relevant observable is the operator representing the density distribution of the atoms.
We prove that the states with definite relative phases are (approximately) the eigenstates of this operator belonging to the eigenvalues (density profiles) exhibiting interference.
According to the projection postulate, the system is projected into one of the eigenstates once the density distribution is measured, resulting in a state with broken symmetry.
Moreover, this simple approach enables us to prove, for the first time to the best of our knowledge, the measurement-induced symmetry breaking for a fermionic system: even if the phases of two superconductors are \textit{a priori} uncertain, the measurement of the current projects the system into a state with a definite relative phase, and a Josephson current flows.

Let us start with the interference of BECs.
We prepare two ideal BECs independently in two separate traps $A$ and $B$ in $D$-dimensional space at zero temperature, each containing exactly $N$ noninteracting atoms.
All the atoms in each trap is condensed in the ground state ($\bm{n}=0$) of the trap, and the state of the couple of BECs is described by the product of the number states 
\begin{equation}
\ket{N,N}=\frac{1}{N!}(\hat{a}_0^\dag)^N(\hat{b}_0^\dag)^N\ket{0},
\label{eqn:NNBEC}
\end{equation}
where $\hat{a}_{\bm{n}}$ and $\hat{b}_{\bm{n}}$ are the bosonic annihilation operators for the $\bm{n}$th levels of traps $A$ and $B$, respectively, satisfying the canonical commutation relations $
[\hat{a}_{\bm{n}},\hat{a}_{\bm{n}'}^\dag]
=[\hat{b}_{\bm{n}},\hat{b}_{\bm{n}'}^\dag]
=\delta_{\bm{n}\bm{n}'}
$,
$
[\hat{a}_{\bm{n}},\hat{a}_{\bm{n}'}]
=[\hat{b}_{\bm{n}},\hat{b}_{\bm{n}'}]
=
[\hat{a}_{\bm{n}},\hat{b}_{\bm{n}'}]
=[\hat{a}_{\bm{n}},\hat{b}_{\bm{n}'}^\dag]
=0
$, and $\ket{0}$ is the vacuum state. 
At time $t=0$, we release the gases from the traps by switching off the trapping potentials, and let them expand in space with no collisions among the atoms.
We then take a photo of the overlapping gases after a time of flight $t$. 
The field operator of the bosonic atoms is expanded as
\begin{equation}
\hat{\psi}(\bm{r},t)
=\sum_{\bm{n}}
\left(
\varphi_{\bm{n}}^{(A)}(\bm{r},t)\hat{a}_{\bm{n}}
+\varphi_{\bm{n}}^{(B)}(\bm{r},t)\hat{b}_{\bm{n}}
\right),
\end{equation}
where $\varphi_{\bm{n}}^{(s)}(\bm{r},t)$ is the wave function evolving from the eigenfunction $\varphi_{\bm{n}}^{(s)}(\bm{r})$ of the $\bm{n}$th level of trap $s\,(=A,B)$, and is a member of a complete set of orthonormal bases, $\int d^D\bm{r}\,\varphi_{\bm{n}}^{(s)*}(\bm{r},t)
\varphi_{\bm{n}'}^{(s')}(\bm{r},t)=\delta_{ss'}\delta_{\bm{n}\bm{n}'}$, $\sum_s\sum_{\bm{n}}\varphi_{\bm{n}}^{(s)}(\bm{r},t)\varphi_{\bm{n}}^{(s)*}(\bm{r}',t)=\delta^D(\bm{r}-\bm{r}')$ \cite{Note:1}.

In the double number state $\ket{N,N}$, the expectation value of the atomic density distribution
\begin{equation}
\hat{\rho}(\bm{r},t)
=\hat{\psi}^\dag(\bm{r},t)\hat{\psi}(\bm{r},t)
\label{eqn:Rho}
\end{equation}
does not exhibit interference, 
\begin{equation}
\langle
\hat{\rho}(\bm{r},t)
\rangle_N
=N\left(
|\varphi_0^{(A)}(\bm{r},t)|^2
+|\varphi_0^{(B)}(\bm{r},t)|^2
\right).
\label{eqn:P1N}
\end{equation}
This is the \textit{independence} of the two BECs, represented by the product state $\ket{N,N}$.
The interference however is actually observed in experiments \cite{ref:InterferenceBEC,Hadzibabic,Schmidmayer}.
Notice that (\ref{eqn:P1N}) is just the expectation value: it predicts the average of all the density profiles observed by many independent and identical experiments.
Although no interference is expected in the average distribution, each \textit{single-shot} photo exhibits interference \cite{JavanainenCiracWalls,ref:CastinDalibard,ReviewsBEC,ref:LeggettQuantumLiquids,ref:BECPethickSmith,IntTh,Boston}.
We need to discuss individual \textit{single} experiments, in spite of the probabilistic nature of quantum mechanics.

Recall here that $\ket{N,N}$ can be expressed as
\begin{equation}
\ket{N,N}
\propto
\int_0^{2\pi}d\theta\,
\ket{\theta},
\label{eqn:NNphase}
\end{equation}
i.e., as a superposition of \textit{phase states} \cite{ref:LeggettSols,ref:CastinDalibard,ref:LeggettQuantumLiquids}
\begin{equation}
\ket{\theta}
=\frac{1}{\sqrt{(2N)!}}
\left(
\frac{
\hat{a}_0^\dag e^{i\theta/2}
+\hat{b}_0^\dag e^{-i\theta/2}
}{\sqrt{2}}
\right)^{2N}
\ket{0}.
\label{eqn:PhaseState}
\end{equation}
In each phase state $\ket{\theta}$, all the $2N$ atoms are in coherent superposition of being in the 0th state of $A$ and being in the 0th state of $B$, with a relative phase $\theta$.
In such a state $\ket{\theta}$, the expectation value of $\hat{\rho}(\bm{r},t)$ exhibits interference
\begin{align}
\langle
\hat{\rho}(\bm{r},t)
\rangle_\theta
&=N\left|
\varphi_0^{(A)}(\bm{r},t)
e^{i\theta/2}
+\varphi_0^{(B)}(\bm{r},t)
e^{-i\theta/2}
\right|^2
\nonumber\\
&\equiv
2N|\Psi_\theta(\bm{r},t)|^2.
\label{eqn:P1Phase}
\end{align}
Equation (\ref{eqn:NNphase}) represents the phase-number uncertainty relation.
We are going to argue that, even if the BECs are prepared in $\ket{N,N}$, a phase state $\ket{\theta}$ is selected from the superposition (\ref{eqn:NNphase}) by measurement, i.e., by taking a photo of the overlapping atomic clouds.
Note that for large $N$ the phase states with different relative phases are approximately orthogonal to each other,
\begin{equation}
\bracket{\theta}{\theta'}
=\left(
\cos\frac{\theta-\theta'}{2}
\right)^{2N}
\xrightarrow{N\to\infty}
\begin{cases}
\medskip
1&(\theta=\theta'),\\
0&(\theta\neq\theta').
\end{cases}
\label{eqn:OrthoPhase}
\end{equation}

The idea is simply to apply the projection postulate in quantum mechanics: the state of a quantum system collapses to the eigenstate of an observable belonging to the measured eigenvalue.
In the present context, the observable is the atomic density distribution operator $\hat{\rho}(\bm{r},t)$ in (\ref{eqn:Rho}), and its eigenvalues represent possible density profiles.
We are going to show that the phase states $\ket{\theta}$ are (approximately) the eigenstates of the observable $\hat{\rho}(\bm{r},t)$.

Notice however that $\hat{\rho}(\bm{r},t)$ does not admit a normalizable eigenstate.
Indeed, when the number of particles in the system is fixed, in the present case $2N$, 
the operator $\hat{\rho}(\bm{r},t)$ in (\ref{eqn:Rho}) in field theory is equivalent to 
\begin{equation}
\hat{\rho}(\bm{r},t)
=\sum_{i=1}^{2N}\delta^D\bm{(}\bm{r}-\hat{\bm{r}}_i(t)\bm{)}
\label{eqn:RhoDelta}
\end{equation}
in quantum mechanics of $2N$ particles \cite{ref:FetterWalecka}.
There is no normalizable eigenstate of the position operator $\hat{\bm{r}}_i(t)$. 
Nonetheless, we will see that the phase states $\ket{\theta}$ are \textit{approximately} the eigenstates of $\hat{\rho}(\bm{r},t)$, when the number of atoms $2N$ is large.

Let us explicitly see the action of $\hat{\rho}(\bm{r},t)$ on $\ket{\theta}$,
\begin{equation}
\hat{\rho}(\bm{r},t)\ket{\theta}
=2N|\Psi_\theta(\bm{r},t)|^2\ket{\theta}+\ket{\delta}.
\label{eqn:AppEigen}
\end{equation}
The contribution $\ket{\delta}$, which is orthogonal to $\ket{\theta}$, is negligibly small.
Indeed, its norm is estimated to be
\begin{equation}
\bracket{\delta}{\delta}(\Delta V)^2
=[\Delta\hat{\rho}(\bm{r},t)]_\theta^2
(\Delta V)^2
\simeq
2N|\Psi_\theta(\bm{r},t)|^2
\Delta V,
\label{eqn:Variance}
\end{equation}
where $[\Delta\hat{\rho}(\bm{r},t)]_\theta^2$ is the variance of $\hat{\rho}(\bm{r},t)$ in $\ket{\theta}$, and we have smeared $\hat{\rho}(\bm{r},t)$ over a small volume $\Delta V$ around each point $\bm{r}$ \cite{Note:2}. 
Equation (\ref{eqn:Variance}) clarifies that $\ket{\delta}$ is $O(\sqrt{N})$ and is negligible compared with the main contribution of $O(N)$ in (\ref{eqn:AppEigen}), when $N$ is large \cite{Note:3}. 
Therefore, (\ref{eqn:AppEigen}) shows that each phase state $\ket{\theta}$ is approximately an eigenstate of $\hat{\rho}(\bm{r},t)$ belonging to an eigenvalue $2N|\Psi_\theta(\bm{r},t)|^2$.

If we think of a superposition of the phase states $\ket{\theta}$, 
\begin{equation}
\ket{\Phi}=\int d\theta\,f(\theta)\ket{\theta},
\label{eqn:superposition}
\end{equation}
the variance of $\hat{\rho}(\bm{r},t)$ is evaluated to be
\begin{align}
&[\Delta\hat{\rho}(\bm{r},t)]_\Phi^2(\Delta V)^2
\simeq
2N^2
\Bigl[
|\chi^2(\bm{r},t)|^2
\,\Bigl(
1-\left|
\overline{e^{i\theta}}
\right|^2
\Bigr)
\nonumber\\
&\quad\ \ \,%
{}+\Re\Bigr\{
\chi^2(\bm{r},t)
\Bigl(
\overline{e^{2i\theta}}
-\overline{e^{i\theta}}^2
\Bigr)
\Bigr\}
\Bigr](\Delta V)^2
+O(N),
\end{align}
where $\chi(\bm{r},t)=\varphi_0^{(B)*}(\bm{r},t)\varphi_0^{(A)}(\bm{r},t)$, $\overline{\,\bullet\,\vphantom{f}}=\int d\theta\,|f(\theta)|^2\,\bullet\,$, and we have used the orthogonality (\ref{eqn:OrthoPhase}).
It is $O(N^2)$ and is not negligibly small, in contrast to (\ref{eqn:Variance}).
Therefore, superpositions (\ref{eqn:superposition}) cannot be (approximate) eigenstates of $\hat{\rho}(\bm{r},t)$, and are not observed in experiments.

In summary, even if the gases are in superposition (\ref{eqn:NNphase}), an interference pattern $2N|\Psi_\theta(\bm{r},t)|^2$ is observed on a \textit{single-shot} photo as an eigenvalue of $\hat{\rho}(\bm{r},t)$, and the system collapses to the corresponding eigenstate $\ket{\theta}$.
The state breaking the $\text{U}(1)$ symmetry is realized by the measurement.
One does not know which phase $\theta$ is selected until the density profile $\hat{\rho}(\bm{r},t)$ is observed.
All the phases are equally probable, and the interference pattern shifts from run to run.
If all the observed patterns are superimposed, we end up with the average profile (\ref{eqn:P1N}) with no fringes, respecting the $\text{U}(1)$ symmetry of the system.

It is worth pointing out that the negligibly small fluctuation (\ref{eqn:Variance}) stems from the particular structures of the observable and of the state.
Let us consider a generic system with $N$ modes in a product state
\begin{equation}
\ket{\psi_1}\otimes\cdots\otimes\ket{\psi_N},
\label{eqn:ProductState}
\end{equation}
where the mode state $\ket{\psi_i}$ ($i=1,\ldots,N$) may differ mode by mode in general, and take an observable of the form
\begin{equation}
\hat{\mathcal{O}}=\sum_{i=1}^N\hat{\mathcal{O}}_i,
\label{eqn:SingleModeOp}
\end{equation}
with $\hat{\mathcal{O}}_i$ acting only on the $i$th mode.
The expectation value and the variance of $\hat{\mathcal{O}}$ in the product state (\ref{eqn:ProductState}) are estimated to be $
\langle\hat{\mathcal{O}}\rangle
=\sum_{i=1}^N\bra{\psi_i}\hat{\mathcal{O}}_i\ket{\psi_i}
\sim O(N)
$ and $
(\Delta\mathcal{O})^2
=\sum_{i=1}^N(
\bra{\psi_i}\hat{\mathcal{O}}_i^2\ket{\psi_i}
-\bra{\psi_i}\hat{\mathcal{O}}_i\ket{\psi_i}^2
)
\sim O(N)
$, respectively.
Therefore, for large $N$, the fluctuation of $\hat{\mathcal{O}}$ in the product state (\ref{eqn:ProductState}) is negligibly small, 
\begin{equation}
\Delta\mathcal{O}/\langle\hat{\mathcal{O}}\rangle\sim O(1/\sqrt{N}).
\label{eqn:RelFluc}
\end{equation}
The same scaling is obtained for the observables consisting of operators which act on multiple but a limited number of modes.

The above analysis on the interference of BECs fits in this framework: the relevant operator, the atomic distribution operator $\hat{\rho}(\bm{r},t)\Delta V$ in (\ref{eqn:RhoDelta}) smeared over $\Delta V$, is actually the sum of the one-body operators, and the phase state $\ket{\theta}$ in (\ref{eqn:PhaseState}) is a product state of the type (\ref{eqn:ProductState}).
That is why the variance scales as in (\ref{eqn:Variance}), and the phase state $\ket{\theta}$ is concluded to be an approximate eigenstate of $\hat{\rho}(\bm{r},t)$.
The superposition of the phase states in (\ref{eqn:superposition}) on the other hand loses the product structure (\ref{eqn:ProductState}), and the variance is not negligibly small anymore.

This observation enables us to go beyond the bosonic systems: the idea of symmetry breaking induced by measurement works also for fermionic systems.
As an example, let us discuss the Josephson effect \cite{ref:Josephson,ref:LeggettQuantumLiquids,ref:Tinkham}.
When two superconductors are connected by a weak link with a thin insulating barrier (Josephson junction), current flows 
from one superconductor to the other, even in the absence of the biases in the chemical potential and in the temperature between the two superconductors.
The current $J_\theta$ depends on the phase difference $\theta=\theta_A-\theta_B$ between the two superconductors $A$ and $B$,
\begin{equation}
J_\theta=J_S\sin\theta.
\label{eqn:Josephson}
\end{equation}
For this Josephson current, the phase difference $\theta$ should be well defined.
We are going to argue that such a state with broken symmetry is realized by measurement even if the system is originally in a superposition state
respecting the $\text{U}(1)$ symmetry: the current measurement selects a relative phase $\theta$, and the Josephson current $J_\theta$ flows.

For concreteness, let us set up a Hamiltonian \cite{ref:Ambegaokar1963,ref:Mahan}
\begin{equation}
\hat{H}=\hat{H}_0+\hat{V},\qquad
\hat{H}_0=\hat{H}_A+\hat{H}_B.
\end{equation}
Here, $\hat{H}_{A}$ and $\hat{H}_{B}$ represent the standard many-body Hamiltonians for the two superconductors $A$ and $B$, respectively, with weak attractive interactions among the electrons giving rise to the superconductivity \cite{ref:Tinkham,ref:FetterWalecka}.
No mean-field approximation is applied here, and the $\text{U}(1)$ symmetry is retained.
The other Hamiltonian 
\begin{equation}
\hat{V}
=\sum_{\bm{k},\bm{k}'}\sum_\sigma
(
T_{\bm{k}\bm{k}'}
\hat{a}_{\bm{k}\sigma}^\dag
\hat{b}_{\bm{k}'\sigma}
+T_{\bm{k}\bm{k}'}^*
\hat{b}_{\bm{k}'\sigma}^\dag
\hat{a}_{\bm{k}\sigma}
)
\end{equation}
provokes the tunneling of electrons through the junction, with $\hat{a}_{\bm{k}\sigma}$ and $\hat{b}_{\bm{k}\sigma}$ being the fermionic annihilation operators for the electrons with wave vector $\bm{k}$ and spin $\sigma\,(={\uparrow},{\downarrow})$ in $A$ and $B$, respectively, satisfying the canonical anticommutation relations $\{\hat{a}_{\bm{k}\sigma},\hat{a}_{\bm{k}'\sigma'}^\dag\}
=\{\hat{b}_{\bm{k}\sigma},\hat{b}_{\bm{k}'\sigma'}^\dag\}
=\delta_{\bm{k}\bm{k}'}\delta_{\sigma\sigma'}
$,
$
\{\hat{a}_{\bm{k}\sigma},\hat{a}_{\bm{k}'\sigma'}\}
=\{\hat{b}_{\bm{k}\sigma},\hat{b}_{\bm{k}'\sigma'}\}
=\{\hat{a}_{\bm{k}\sigma},\hat{b}_{\bm{k}'\sigma'}\}
=\{\hat{a}_{\bm{k}\sigma},\hat{b}_{\bm{k}'\sigma'}^\dag\}
=0
$.
The matrix elements $T_{\bm{k}\bm{k}'}$ characterize the tunneling process, which is assumed to be symmetric, $T_{\bm{k}\bm{k}'}=T_{(-\bm{k}')(-\bm{k})}^*$. 
The current operator $\hat{J}$, the relevant operator for the present argument, is then defined by 
\begin{equation}
\hat{J}
=-ie[\hat{N}_A,\hat{H}]
=
2e\Im
\sum_{\bm{k},\bm{k}'}
\sum_\sigma
T_{\bm{k}\bm{k}'}
\hat{a}_{\bm{k}\sigma}^\dag
\hat{b}_{\bm{k}'\sigma},
\label{eqn:I}
\end{equation}
where $\hat{N}_A=\sum_{\bm{k}}\sum_\sigma\hat{a}_{\bm{k}\sigma}^\dag\hat{a}_{\bm{k}\sigma}$ ($\hat{N}_B=\sum_{\bm{k}}\sum_\sigma\hat{b}_{\bm{k}\sigma}^\dag\hat{b}_{\bm{k}\sigma}$) is the number of electrons in $A(B)$ \cite{ref:Ambegaokar1963,ref:Mahan}.
Note that $\hat{N}_{A(B)}$ commutes with $\hat{H}_{A(B)}$, and the total number of electrons $\hat{N}_A+\hat{N}_B$ is preserved by $\hat{H}$, thanks to the $\text{U}(1)$ symmetry of the Hamiltonians.

Suppose now that the two superconductors are given at zero temperature in a superposition state
\begin{equation}
\ket{\Xi}
=\int d\theta_A
\int d\theta_B\,
c_{\theta_A,\theta_B}\ket{\Omega_{\theta_A},\Omega_{\theta_B}},
\label{eqn:SuperBCS}
\end{equation}
e.g., in a state respecting the $\text{U}(1)$ symmetry, where
\begin{align}
\ket{\Omega_{\theta_A},\Omega_{\theta_B}}
={}&
\prod_{\bm{k}}(u_k+v_ke^{i\theta_A}\hat{a}_{\bm{k}{\uparrow}}^\dag \hat{a}_{-\bm{k}{\downarrow}}^\dag)
\nonumber\\
&{}\times
\prod_{\bm{k}'}(u_{k'}+v_{k'}e^{i\theta_B}\hat{b}_{\bm{k}'{\uparrow}}^\dag \hat{b}_{-\bm{k}'{\downarrow}}^\dag)
\ket{0}
\label{eqn:DoubleBCS}
\end{align}
is the product of the standard BCS states \cite{ref:Tinkham}.
We assume that $u_k$ and $v_k$ are common for both superconductors.
We then let the system evolve by the Hamiltonian, and see the current $\hat{J}$ in the stationary limit.
The stationary state, which is a \textit{nonequilibrium steady state} (NESS) \cite{NESS}, in which the current flows steadily, is mathematically generated by the M\o ller wave operator \cite{ref:ScatteringTaylor} 
\begin{equation}
\hat{W}=\lim_{t\to\infty}e^{-i\hat{H}t/\hbar}e^{i\hat{H}_0t/\hbar}.
\end{equation}
We therefore look at the ``scattered'' current operator
\begin{equation}
\hat{J}_\infty
=\hat{W}^\dag
\hat{J}
\hat{W}.
\label{eqn:It}
\end{equation}
We are going to show that each double BCS state $\ket{\Omega_{\theta_A},\Omega_{\theta_B}}$ in the superposition (\ref{eqn:SuperBCS}) is approximately an eigenstate of $\hat{J}_\infty$ belonging to the eigenvalue (\ref{eqn:Josephson}).

Notice first that the BCS states with different phases are approximately orthogonal to each other,
\begin{equation}
\bracket{\Omega_\theta}{\Omega_{\theta'}}
=e^{\sum_{\bm{k}}\ln(u_k^2+v_k^2e^{-i(\theta-\theta')})}
\xrightarrow{V\to\infty}
\begin{cases}
\medskip
1&(\theta=\theta'),\\
0&(\theta\neq\theta').
\end{cases}
\end{equation}
It remains the case even if a bounded operator is inserted.
In particular, for $(\theta_A,\theta_B)\neq(\theta_A',\theta_B')$,
\begin{equation}
\bra{\Omega_{\theta_A},\Omega_{\theta_B}}
\hat{J}_\infty
\ket{\Omega_{\theta_A'},\Omega_{\theta_B'}}
\xrightarrow{V\to\infty}
0.
\end{equation}
This means that $\hat{J}_\infty$ is diagonal within the subspace spanned by $\{\ket{\Omega_{\theta_A},\Omega_{\theta_B}}\}$.
This however is not enough to conclude that each $\ket{\Omega_{\theta_A},\Omega_{\theta_B}}$ is an eigenstate of $\hat{J}_\infty$, since the
action of $\hat{J}_\infty$ on $\ket{\Omega_{\theta_A},\Omega_{\theta_B}}$ might give rise to a component $\ket{\delta}$ perpendicular to this subspace as in (\ref{eqn:AppEigen}), 
\begin{equation}
\hat{J}_\infty\ket{\Omega_{\theta_A},\Omega_{\theta_B}}
=J_\theta\ket{\Omega_{\theta_A},\Omega_{\theta_B}}
+\ket{\delta}.
\label{eqn:DeltaBCS}
\end{equation}
Still, $\ket{\delta}$ turns out to be negligibly small, by estimating $\bracket{\delta}{\delta}$, i.e., the variance of $\hat{J}_\infty$ in the state $\ket{\Omega_{\theta_A},\Omega_{\theta_B}}$, as we did in (\ref{eqn:Variance}) for the bosonic case.
To this end, it suffices to check the structures of the current operator $\hat{J}_\infty$ and of the state $\ket{\Omega_{\theta_A},\Omega_{\theta_B}}$.
The former is shown, in the standard mean-field approximation within each single sector $\ket{\Omega_{\theta_A},\Omega_{\theta_B}}$ \cite{ref:Tinkham,ref:FetterWalecka}, to be composed of two-mode operators $\hat{J}_{\bm{k}\bm{k}'}$ acting on modes $\bm{k}$ and $\bm{k}'$ \cite{Note:4}, 
\begin{equation}
\hat{J}_\infty\simeq \sum_{\bm{k},\bm{k}'}\hat{J}_{\bm{k}\bm{k}'},
\end{equation}
while the latter, defined in (\ref{eqn:DoubleBCS}), is a product state of different modes of the type (\ref{eqn:ProductState}).
Therefore, by the theorem presented around (\ref{eqn:ProductState})--(\ref{eqn:RelFluc}), the fluctuation of $\hat{J}_\infty$ in each $\ket{\Omega_{\theta_A},\Omega_{\theta_B}}$ is negligibly small in the macroscopic limit, and $\ket{\delta}$ in (\ref{eqn:DeltaBCS}) is proved to be negligibly small.
That is, each double BCS state $\ket{\Omega_{\theta_A},\Omega_{\theta_B}}$ is approximately an eigenstate of $\hat{J}_\infty$.
The corresponding eigenvalue is given by $J_\theta=\bra{\Omega_{\theta_A},\Omega_{\theta_B}}
\hat{J}_\infty
\ket{\Omega_{\theta_A},\Omega_{\theta_B}}$, which in the standard mean-field approximation within the sector $\ket{\Omega_{\theta_A},\Omega_{\theta_B}}$ reproduces the result by Ambegaokar and Baratoff \cite{ref:Ambegaokar1963}, yielding the Josephson current (\ref{eqn:Josephson}).

We then have the following scenario.
Once the current $\hat{J}_\infty$ is measured in the NESS, one of its eigenvalues $J_\theta$ is observed, and the state of the couple of superconductors collapses.
Since the eigenstates $\ket{\Omega_{\theta_A},\Omega_{\theta_B}}$ with a common relative phase $\theta=\theta_A-\theta_B$ are all degenerated in the same eigenvalue $J_\theta$, the system is projected onto the subspace formed by them,
\begin{equation}
\ket{\Xi}
\to
\left.
\int d\bar{\theta}\,
c_{\theta_A,\theta_B}\ket{\Omega_{\theta_A},\Omega_{\theta_B}}
\right|_{\theta_A-\theta_B=\theta},
\label{eqn:BCSRel}
\end{equation}
where $\bar{\theta}=(\theta_A+\theta_B)/2$ remains unfixed, but the relative phase $\theta$ is fixed.
The state (\ref{eqn:BCSRel}) breaking the $\text{U}(1)$ symmetry is thus realized by the measurement, and the Josephson current $J_\theta$ in (\ref{eqn:Josephson}) flows.

The above analysis on the BEC interference shows that an interference pattern is \textit{certainly} observed in each single experiment, regardless of its unpredictable spatial offset, while the density profiles with no interference are rarely found.
In this sense, the density profiles exhibiting interference are \textit{typical} among all possible configurations of the atoms, and so are the corresponding states breaking the $\text{U}(1)$ symmetry.
We have shown that it is the case also for the BCS states for fermions and proved that the Josephson effect arises when the states with a definite relative phase are selected by a current measurement.
Recently, there are intensive attempts to replace the equal \textit{a priori} probability postulate of statistical mechanics with \textit{typicality} \cite{Typicality}: 
a typical pure state randomly sampled from an energy shell in the Hilbert space well represents the microcanonical ensemble.
In the presence of condensation, however, such naive sampling from an energy shell does not work, since in that way we might typically pick a superposition of different phases, which is not observed in real experiments.
Our results suggest that we need to care what is measured, i.e., the relevant observable and its spectrum, to properly sample a state realized in each \textit{single} event.
This subject deserves detailed study, to generalize the typicality approach to statistical mechanics in the presence of condensation.

This work is supported by Grants-in-Aid for Young Scientists (B) (No.\ 26800206) and for Scientific Research (C) (No.\ 26400406) from JSPS, Japan, and by Waseda University Grants for Special Research Projects (2013A-982 and 2013B-147).


\begin{thebibliography}{10}
\bibitem{Note:2}
This is necessary to avoid $\delta^D(0)$ originating from the completeness of
  the set of wave functions $\{\varphi_{\bm{n}}^{(s)}(\bm{r},t)\}$. What is relevant is not the
  density but the number of atoms. Such smearing is actually done in real
  experiments: photos have finite resolutions.
\bibitem{Note:3}
We suppose that $N|\Psi_\theta(\bm{r},t)|^2\Delta V$ is large where
  $\Psi_\theta(\bm{r},t)$ is nonvanishing: a sufficiently large
  number of atoms should be included in $\Delta V$.
\bibitem{Note:4}
The current operator $\hat{J}$ in (\ref{eqn:I}) is
  given by the sum of the two-mode operators. This structure is preserved by
  the scattering by $\hat{W}$, since in the
  mean-field approximation the Hamiltonians are quadratic.
\end{thebibliography}

\end{document}